\pdfoutput=1
\documentclass[aps,prl,reprint,groupedaddress,twocolumn,superscriptaddress]{revtex4-1}
\usepackage{graphicx}
\usepackage{hyperref}
\begin{document}
\title{
Optical Traps for sympathetic Cooling of Ions with ultracold neutral Atoms
}
\author{J. Schmidt}
\address{Laboratoire Kastler Brossel, UPMC-Sorbonne Universités, CNRS, ENS-PSL Research University, Collège de France, 4 place Jussieu, Paris 75005, France}
\address{Albert-Ludwigs-Universität Freiburg, Physikalisches Institut, Hermann-Herder-Straße 3, 79104 Freiburg, Germany}
\author{P. Weckesser}
\author{F. Thielemann}
\author{T. Schaetz}
\address{Albert-Ludwigs-Universität Freiburg, Physikalisches Institut, Hermann-Herder-Straße 3, 79104 Freiburg, Germany}
\author{L. Karpa}
\email[]{leon.karpa@physik.uni-freiburg.de}
\address{Albert-Ludwigs-Universität Freiburg, Physikalisches Institut, Hermann-Herder-Straße 3, 79104 Freiburg, Germany}
\date{\today}
\begin{abstract}
We report the trapping of ultracold neutral $ \text{Rb}$ atoms and $ \text{Ba}^+ $ ions in a common optical potential in absence of any radiofrequency (RF) fields. We prepare $ \text{Ba}^+ $ at $ 370 ~ \mu K $ and demonstrate efficient sympathetic cooling by $100 ~ \mu K $ after one collision. Our approach is currently limited by the $ \text{Rb}$ density and related three-body losses, but it overcomes the fundamental limitation in RF traps set by RF-driven, micromotion-induced heating. It is applicable to a wide range of ion-atom species, and may enable novel ultracold chemistry experiments and complex many-body dynamics.
\end{abstract}
\maketitle
Ultracold ensembles of ions and atoms have been attracting broad interest across disciplines for more than a decade\,\cite{Smith2005,Grier2009,Schmid2010,Zipkes2010,Ravi2012,Hall2012,Rellergert2011,Haerter2014,Haze2018,Meir2016,Tomza2019,Deiglmayr2012}. This is largely owed to the prospects of reaching a temperature regime where phenomena are governed by quantum mechanics, which allows new insights into fundamental physics, as well as control over interactions and chemical reactions. The long-range interaction arising from the polarization of atoms opens up novel pathways to access and control many interesting effects and phenomena, such as Feshbach resonances\,\cite{Idziaszek2009,Tomza2015}, the formation of novel molecular states \cite{Cote2002,Schurer2017},  as well as applications in quantum information processing\,\cite{Secker2016,Doerk2010} and simulations\,\cite{Bissbort2013}. Reaching ultralow temperatures for both species is favourable, and in most cases a prerequisite\,\cite{Krych2011,Meir2016,Fuerst2018}. Experiments in hybrid traps, confining ions with RF and atoms with optical fields, have shown that sympathetic cooling is highly efficient in the millikelvin energy range and above\,\cite{Zipkes2010,Ravi2012,Haze2018,Sivarajah2012,Haerter2014,Tomza2019}. However, it was observed that the presence of RF fields inevitably invokes micromotion-induced heating\,\cite{Cetina2012,Fuerst2018,Grier2009,Meir2016}. The latter is inherent to all RF traps and has precluded access to ultralow collision energies. More recently, promising strategies for specific scenarios have been developed, requiring e.g. Rydberg atoms and ions\,\cite{Secker2016}, extremely large ion-atom mass ratios\,\cite{Cetina2012,Fuerst2018,Feldker2019} or Rydberg excitations within homonuclear atomic ensembles\,\cite{Kleinbach2018}. Yet a universal approach for reaching deep into the quantum regime of interaction for generic combinations of atom and ion species, or even (higher-dimensional) Coulomb crystals, remains an outstanding challenge.

In this Letter, we demonstrate a generic method based on optical ion trapping\,\cite{Schneider2010,Lambrecht2017,Schmidt2018,Schaetz2017,Karpa2019,Schneider2012}, completely overcoming the limitation imposed by micromotion-induced heating\,\cite{Cetina2012}. We use bichromatic optical trapping potentials to simultaneously trap and control ions and neutral atoms. We observe highly efficient sympathetic cooling of single Doppler-cooled $ ^{138}\text{Ba}^+ $ ions when immersed in a cloud of ultracold $^{87}\text{Rb}$ atoms. We further demonstrate methods for effectively isolating $ \text{Ba}^+ $ from parasitic ions. Our results pave the way towards realizations of ion-atom collision experiments in the quantum dominated regime.

\begin{figure}[h!!!]
\centering
\includegraphics[width = 0.45 \textwidth]{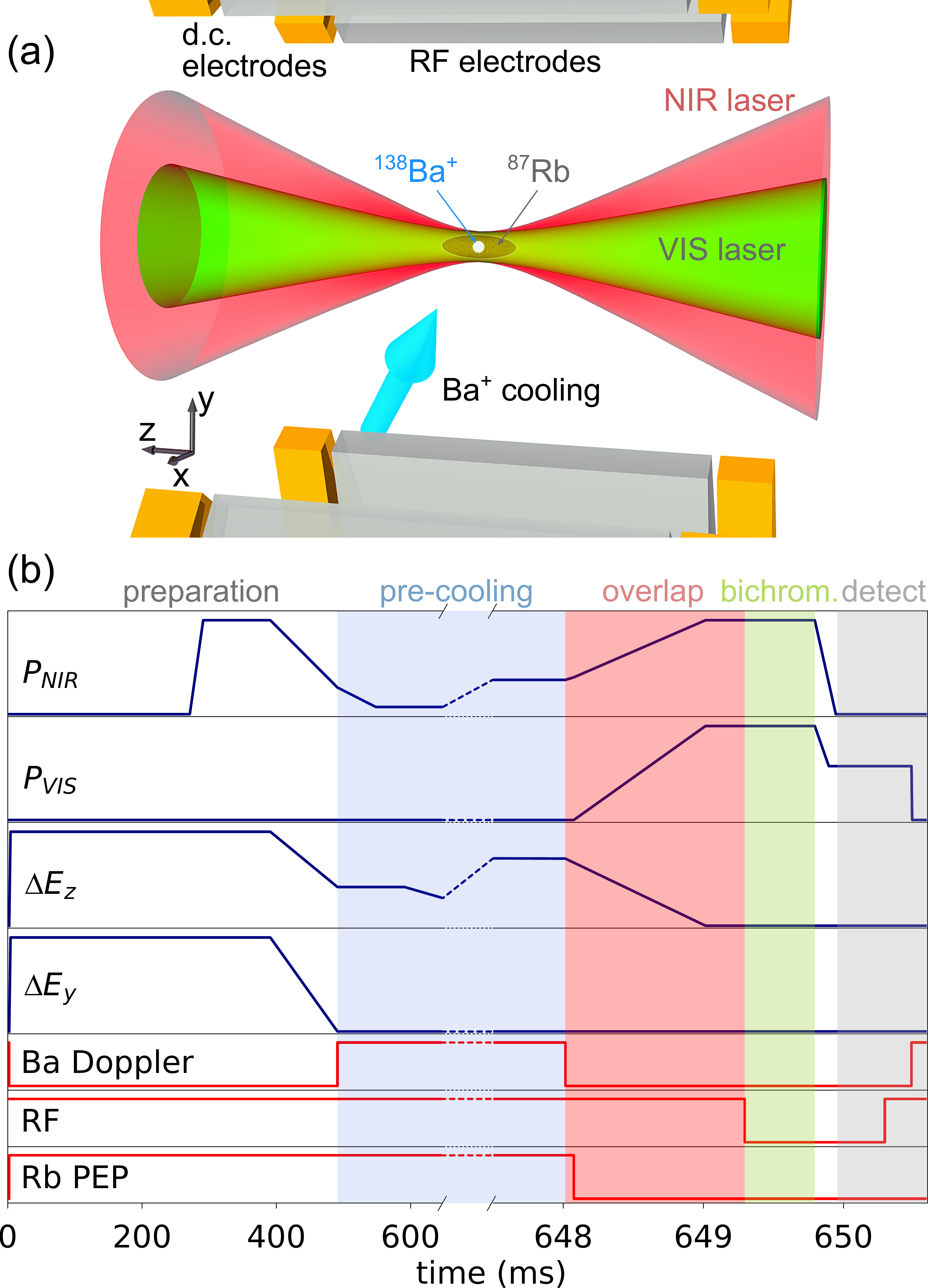}
\caption{
 Schematic of the experimental setup (not to scale) and protocol. \textbf{(a)} Bichromatic dipole traps (biODT) comprised of two lasers, VIS and NIR are used to simultaneously trap a $ ^{138}\text{Ba}^+ $ ion and a cloud of ultracold $^{87}\text{Rb}$ atoms. The linear Paul trap (ion-electrode distance: $ 9 ~ \text{mm} $) is only employed for preparation and detection of the laser cooled ion (d.c. electrodes for axial confinement, compensation and offset fields).
 \textbf{(b)} Simplified experimental sequence showing the most relevant parameters (scales magnified on the right): $ P_{NIR}, P_{VIS}, \Delta E_z, \Delta E_y  $ denote the optical powers of the dipole beams and the electric offset fields in the $ z $ and $ y $ directions. 3 lowermost traces: $ \text{Ba}^+ $ laser cooling (Ba Doppler), RF amplitude (RF), and parametric excitation pulses (PEP). The sequence comprises a preparation (left unshaded), pre-cooling (blue shaded), overlapping (red), interaction (green), and detection phase (grey). 
 During the latter, we detect the survival of $ \text{Ba}^+ $ after trapping in the VIS beam only.
 Loading the $ \text{Rb}$ MOT, transfer into a crossed NIR dipole trap and evaporation therein are not depicted.
 }
\label{fig:schematic}
\end{figure}
We adapt the experimental setup described in\,\cite{Lambrecht2017,Schmidt2018} to allow for optical trapping of ions and atoms as shown in Fig.\ref{fig:schematic}(a). %
We first initialize the $ \text{Ba}^+ $ by photoionizing $ \text{Ba} $ atoms emitted from an oven and trapping the ions in a linear Paul trap (RF trap). It features radial and axial secular frequencies of $ \omega_{(r,ax)}^{\text{Ba}^+}/(2 \pi) \approx (100 , 12.5) ~ \text{kHz} $, where the d.c. electrodes provide the axial confinement. Subsequently, the ions are laser cooled to the Doppler limit of $ T_D^{\text{Ba}^+} \approx 370 ~ \mu \text{K} $. We compensate stray electric fields to the level of $ E_{str} \le 7 \, \text{mV} \, \text{m}^{-1} $\,\cite{Huber2014,Karpa2019}, and align two axially overlapped counter-propagating dipole trap laser beams (VIS and NIR) with a $ \text{Ba}^+ $ confined at the center of the RF trap $ (z = 0 ~ \text{m}) $\,\cite{Lambrecht2017,Schmidt2018}. The two beams are operated at wavelengths $ \lambda_{VIS} = 532 \, \text{nm} $ and $ \lambda_{NIR} = 1064 \, \text{nm} $ while the beam waist radii of $ w_0^{NIR} \approx w_0^{VIS} = 3.7 \pm 0.05 ~ \mu \text{m}$  are approximately matched at $ z=0 $.

\begin{figure}[!ht]
\includegraphics[width = 0.49 \textwidth]{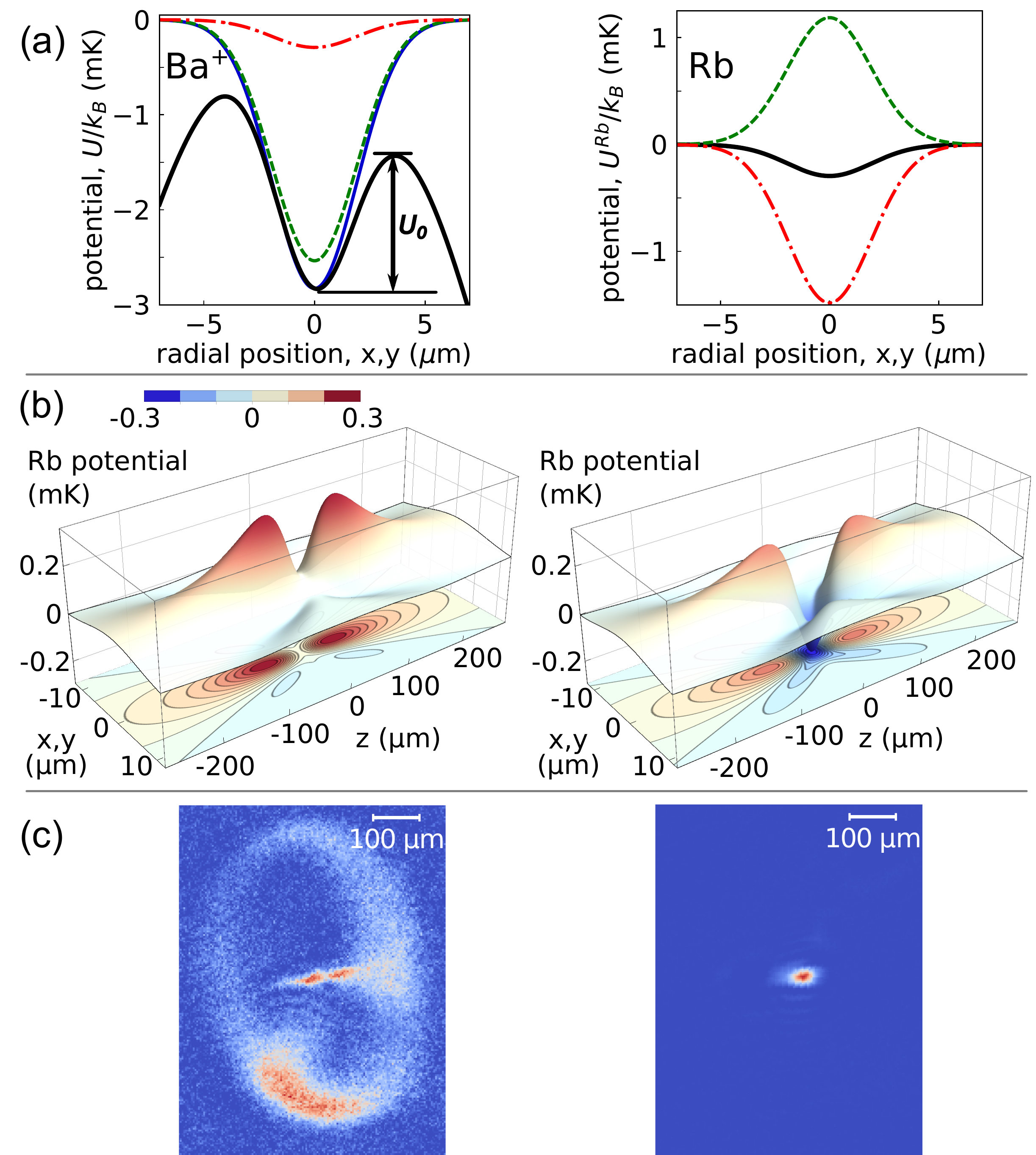}
\caption{
 Bichromatic trapping potentials for $ \text{Ba}^+ $ and $ \text{Rb}$. \textbf{(a)} (left) Calculated optical potential for $ \text{Ba}^+ $ at $ z = 0 $ (blue, solid) with VIS (green, dashed) and NIR (red, dash-dotted) contributions. Taking into account electrostatic defocusing and stray fields (black, thick) yields an effective trap depth of $ U_0 / k_B \approx 1.4 ~ \text{mK} $.  (right) Calculated bichromatic potential for $ \text{Rb}$ (solid, black) with the repulsive VIS and attractive NIR potentials adjusted to provide an effective trap depth of $ U_0^{\text{Rb}} = U_{NIR}^{\text{Rb}}(x,y=0) + U_{VIS}^{\text{Rb}}(x,y=0) \approx k_B \times 300 ~ \mu \text{K} $ within the $ z = 0 $ plane.
 \textbf{(b)} Corresponding bichromatic potential landscape for $ \text{Rb}$ (contour plot at the bottom). (left) In the unbalanced case $ \left| U_{VIS}^{\text{Rb}} \right| >\left| U_{NIR}^{\text{Rb}} \right| $: the potential at $ z = 0 $ is rendered repulsive featuring shallow lobes around $ z \approx \pm 50 ~ \mu \text{m}$. (right) In the balanced case $ U_0^{\text{Rb}} / k_B \approx 300 ~ \mu \text{K} $, the global minimum is located at the center.    
 \textbf{(c)} Absorption images of $ \text{Rb}$ ($ N \approx 4 \times 10^{3} $ atoms) in a bichromatic potential after $ \sim 100 ~ \mu \text{s} $ time-of-flight taken nearly from the $ x $ direction (projecting into the $y-z$ plane): (left) VIS ramp is leading with respect to the NIR ramp and (right) with appropriately synchronized ramp-up of $ P_{VIS} $ and $ P_{NIR} $. In the unbalanced case, atoms have been ejected from the center, while some atoms still accumulate at the local minima around $ z \approx \pm 50 ~ \mu \text{m}$.
 }
\label{fig:bichro}

\end{figure}

The sequence for studying atom-ion interactions comprises four distinct stages shown in Fig.\ref{fig:schematic}(b): preparation, pre-cooling, overlapping and bichromatic phase. The latter ensures transfer of the overlapped ion-atom system from the hybrid trap (RF for $ \text{Ba}^+ $, NIR and VIS for $ \text{Rb}$) to the bichromatic optical dipole traps (biODT) where we turn off the RF fields. In the final detection phase, the survival of $ \text{Ba}^+ $ in the VIS trap is determined after its transfer back to the RF trap via fluorescence imaging on a charge-coupled device (CCD) camera (see Fig.\ref{fig:schematic}(b)). Repeating this experiment yields the trapping probability $ p_{opt} $. A variation of the trap depth $ U_0 $ (see Fig. \ref{fig:bichro}(a), left) allows to derive the $ \text{Ba}^+ $ temperature, $ T^{\text{Ba}^+} $\,\cite{Schneider2010,Lambrecht2017,Schaetz2017,Schmidt2018,Karpa2019}.%

We start the preparation phase by loading $ \text{Rb}$ into a magneto-optical trap (MOT) while spatially separating the initialized $ \text{Ba}^+ $ from the atom cloud. Here, we add electric offset fields $\Delta E_y \approx 15  ~ \text{V} ~ \text{m}^{-1}$ and $ \Delta E_z \approx 0.8 ~ \text{V} ~ \text{m}^{-1}$ relative to the configuration determined during the stray field compensation. The former offset the ion radially (axially) by $\Delta y \approx 30 ~ \mu \text{m} $ $( \Delta z \approx 300 \, \mu \text{m} )$. After loading the MOT for about $ \Delta t_{MOT} \approx 250 ~ \text{ms} $, we first transfer the $ \text{Rb}$ into a crossed NIR optical trap (not shown in Fig.\ref{fig:schematic}), and after $ 50 ~ \text{ms} $ into a focused NIR dipole trap. Forced evaporation therein yields a cloud of $\sim 10^3 $ atoms.%

At the end of the preparation phase, we ramp $ \Delta E_y $ to zero and adjust $ \Delta E_z $ to position $ \text{Ba}^+ $ about $ 100 ~ \mu \text{m} $ away from the trap center ($ z=0 $). There, it is Doppler cooled for approximately $ 150 \, \text{ms} $ (pre-cooling phase in Fig.\ref{fig:schematic}(b)). Subsequently, during the ion-atom overlapping phase in the hybrid trap, the VIS laser is ramped up within $ \tau_{ramp} = 1 \, \text{ms} $ to $ P_{VIS} \approx 470 ~ \text{mW} $. This corresponds to an effective trap depth of $ U_0 / k_B \approx 1.4 ~ \text{mK} $ for $ \text{Ba}^+ $ (see Fig.\ref{fig:bichro}(a), left), where $ k_B $ is the Boltzmann constant. Simultaneously, $ P_{NIR} $ has to be increased from $ 26 $ to $ 65 ~ \text{mW} $ in order to compensate for the repulsive effect of the VIS laser on $\text{Rb}$ (see Fig.\ref{fig:bichro}(a), right). %
This requires maintaining the ratio $ P_{NIR} / P_{VIS} $ within a narrow range to keep the trap depth $ U^{\text{Rb}}_0 $ for $ \text{Rb}$ close to constant (see Fig.\ref{fig:bichro}(b,c), right panels)). A deviation by $ \sim 10 \% $ leads either to substantial heating due to compression by the NIR laser or to atom loss induced by the repulsive VIS laser (see Fig.\ref{fig:bichro}(b,c), left panels)). While building up the biODT, the $ \text{Ba}^+ $ is transported to the center of the $ \text{Rb}$ cloud by ramping $ \Delta E_z $ to zero. At this point, overlap of $ \text{Ba}^+ $ and $ \text{Rb}$ in the hybrid trap is established. Note that the biODT deviates from the idealized assumption of circular Gaussian profiles shown in Fig.\ref{fig:bichro}(a,b). In our current experimental realization both dipole beams are astigmatic, not perfectly Gaussian and prone to drifts as well as intensity and pointing noise. This causes fluctuations of the number of $ \text{Rb}$ atoms, their spatial distribution and affects the control of the local density. We illustrate two extreme scenarios in Fig.\ref{fig:bichro}(c).

The subsequent bichromatic phase allows for observing ion-atom interactions in absence of RF fields. It is initiated by transferring $ \text{Ba}^+ $ from the RF to the biODT by ramping the RF field to zero. The duration of confinement in the biODT amounts to $ \tau \approx 0.5 ~ \text{ms} $. The sequence is concluded by ramping the NIR trap to zero, thereby ejecting $ \text{Rb}$ out of the VIS beam, leaving the ion in the VIS trap. We repeat the sequence to measure $ p_{opt} $ and $ T^{\text{Ba}^+} $. %

\begin{figure}[h]
\includegraphics[width = 0.45 \textwidth]{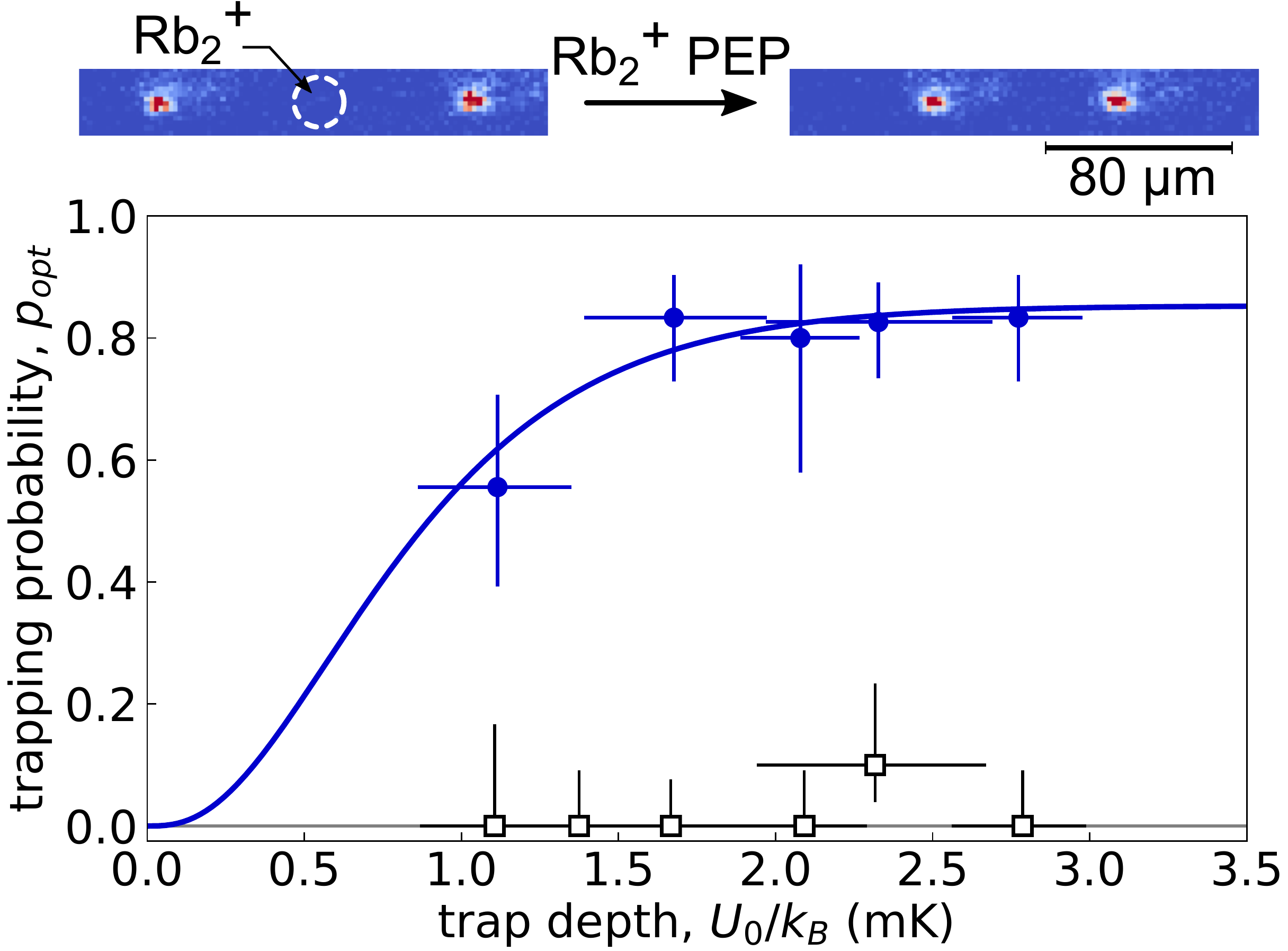}
\caption{
Isolation efficiency of $ \text{Ba}^+ $ from parasitic ions before transfer into the bichromatic trap.
 	Optical trapping probability $ p_{opt} $ in the VIS trap of depth $ U_0 / k_B $ at the end of the overlapping phase without (black hollow squares) and with parametric excitation pulses (PEP) applied during loading and evaporation of $ \text{Rb}$. Error bars: (trap depth) upper bounds of 1 $ \sigma $ uncertainties extracted from bootstrapping; ($ p_{opt} $) $ 1 ~\sigma $ confidence intervals calculated from the underlying binomial distribution. (top) Three-ion chain with a $ \text{Rb}_2^+ $ molecule embedded in a $ \text{Ba}^+ $ crystal before (left) and after (right) application of PEP tuned in resonance with $ \text{Rb}_2^+ $, selectively removing the parasitic ion\,\cite{Haerter2013b}. In our experiment: $ \text{Rb}$ and $ \text{Ba}^+ $ are spatially separated while PEP is active, such that $\text{Rb}^+$ and $\text{Rb}_2^+ $ ions are expelled before crystallizing.
 	 }
\label{fig:bouncers}

\end{figure}

So far, we have neglected parasitic ions, assuming that only $^{138} \text{Ba}^+ $ was present in the RF trap. However, it has been demonstrated that photo-associative processes at $ \lambda_{NIR} $ within the dense neutral $ \text{Rb}$ cloud can lead to the formation of $ \text{Rb}^+ $ and $ \text{Rb}_2^+ $\,\cite{Haerter2013b}. This effect is enhanced due to the still large $ P_{VIS} $ (and accordingly $ P_{NIR} $) required to initially trap the comparatively hot ion at $ T^{\text{Ba}^+}_D$. Both parasitic species fulfil the criteria for stable confinement in our RF trap, as shown for a chain of a single $ \text{Rb}_2^+ $ and two $ \text{Ba}^+ $ ions in Fig.\ref{fig:bouncers} (top left panel). Their disruptive appearance within the cloud leads to heating of $ \text{Ba}^+ $, thereby thwarting its transfer into the shallow biODT. To quantitatively investigate this effect and potential countermeasures, we measure $ p_{opt} $ immediately after establishing overlap of $ \text{Ba}^+ $ in the hybrid trap with a cloud of $ N \approx (1 \pm 0.15)\times 10^3$ atoms at a temperature of $ T^{\text{Rb}} \approx 30 \pm 10 ~ \mu \text{K} $ (open squares in Fig.\ref{fig:bouncers}). We find that $ p_{opt} $, here a measure of transfer efficiency of $ \text{Ba}^+ $ from the hybrid trap to biODT, is consistent with zero for $ U_0/k_B $ up to $ 3 ~ \text{mK} $. This imposes a lower bound on the kinetic energy in the range of several $ k_B \times \text{mK} $. In our case, the parasitic ions have to be selectively removed from the hybrid trap prior to overlapping $ \text{Ba}^+ $ with the now purified $ \text{Rb}$ ensemble. To this end we apply $ \Delta E_z $ and $ \Delta E_y $ to offset $ \text{Ba}^+ $ in combination with a series of chirped parametric excitation pulses (PEP). That is, we modulate the amplitude of the RF field at twice the radial secular frequencies of both the $ \text{Rb}^+ $ and $ \text{Rb}_2^+ $ (sweep of $ \pm 5$, and $\pm 2.5 ~ \text{kHz}$, respectively)\,\cite{Haerter2013b}. As shown in Fig.\ref{fig:bouncers}(top), this method efficiently removes the parasitic ions, even if they are embedded into a $ \text{Ba}^+ $ Coulomb crystal. The measured $ p_{opt} $ reveals an improved transfer and trapping performance to $ p_{opt} \approx 0.8$ (see blue circles in Fig.\ref{fig:bouncers}). We account for $ p_{opt} < 1 $ by considering the finite maximal $ p_{opt} $ in the radial-cutoff model\,\cite{Schneider2012}. A fit to the data with the adapted model yields a temperature of $T_{init}^{\text{Ba}^{+}} = 356 \pm 30 \, \mu \text{K}$. This is consistent with $ T^{\text{Ba}^+}_D $. We conclude that the reduced transfer efficiency is due to loss of $ \text{Ba}^+ $ rather than heating. The observed loss could be caused e.g. by a residual survival probability of motionally excited $ \text{Rb}^+ $ or $ \text{Rb}_2^+ $ on larger orbits within the RF trap or by motional excitation of $ \text{Ba}^+ $ during the transport phases with amplitudes exceeding the effective range of the pre-cooling lasers. Notwithstanding, PEP combined with a laser pre-cooling phase (see Fig.\ref{fig:schematic}(b)) efficiently isolates $ \text{Ba}^+ $ at the end of the overlapping sequence.

We now study the interaction of $ \text{Ba}^+ $, pre-cooled to $ T_D^{\text{Ba}^+} $, and $ \text{Rb}$ within the hybrid trap in dependence on the duration of overlap $ \tau_{hyb} $.
Even for well compensated $ E_{str} $, we observe that $ \text{Ba}^+ $ experiences heating. 
For $ \tau_{hyb} $ on the order of $ \text{ms} $ and atomic densities of $n_0 \sim 10^{12} \, \text{cm}^{-3} $ we measure a heating rate $ R_{RF} = 0.3 \pm 0.14 \, \text{K} \, \text{s}^{-1} $. 
In contrast, in absence of $ \text{Rb}$, $ R_{RF} $ remains below our detection limit.  
We attribute the former to micromotion-induced heating, in agreement with previous results obtained in  hybrid traps\,\cite{Meir2016}.

\begin{figure}[h!!!]
\includegraphics[width = 0.45 \textwidth]{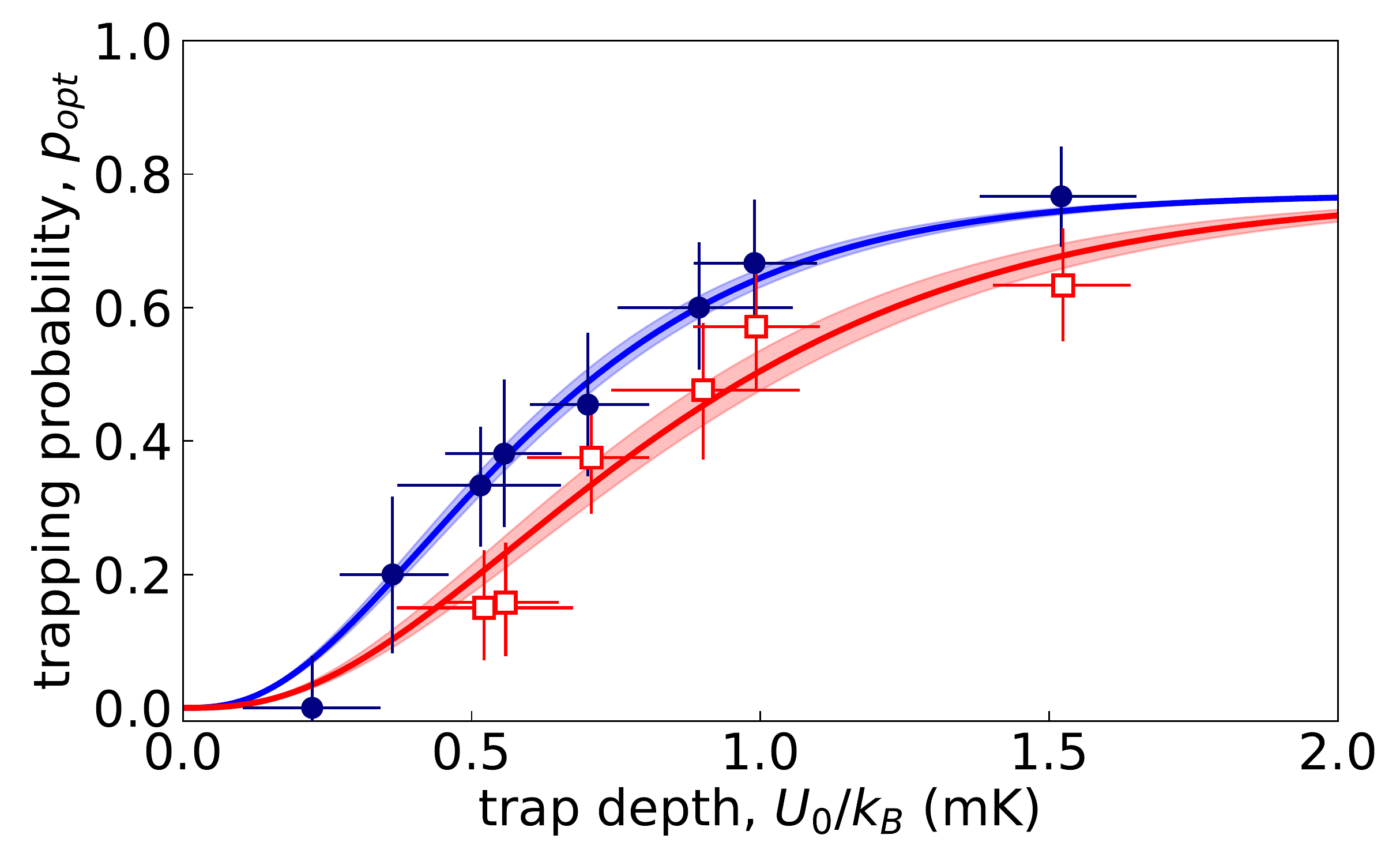}
 	\caption{
Sympathetic cooling of a $ \text{Ba}^+ $ ion in a cloud of ultracold $ \text{Rb}$ atoms. Open squares: experimental data taken after placing the ion in the bichromatic trap in absence of atoms. An orthogonal distance regression fit with a modified radial cutoff model\,\cite{Schneider2012} to the data (lower solid line) yields a temperature of $ T_{init}^{\text{Ba}^+} = 357 \pm 22 \, \mu \text{K} $. Full circles: the same experiment carried out with atoms. A fit to the data (upper solid line) yields $ T_{symp}^{\text{Ba}^+} = 259 \pm 10 \, \mu \text{K} $, showing evidence for sympathetic cooling below the Doppler limit $ T_D^{\text{Ba}^+} $. Shaded regions: bounds corresponding to the fit standard errors. Error bars: (trap depth) upper bounds of 1 $ \sigma $ uncertainties extracted from bootstrapping; ($ p_{opt} $) upper bounds of $ 1 ~\sigma $ confidence intervals calculated from the underlying binomial distribution.
 }
\label{cooling}

\end{figure}

We finally investigate the interaction of $ \text{Ba}^+ $ and $ \text{Rb}$ within the biODT. 
We carry out the complete experimental sequence, now including the bichromatic phase with a duration $ \tau_{int} $. %
To distinguish the impact of the ion-atom interaction from any systematic effects, we either keep the $ \text{Rb}$, or dismiss it directly after transferring $ \text{Ba}^+ $ into the biODT. We start this stage of the sequence by confining $ N \approx 500 \pm 150 $ $ \text{Rb}$ in the biODT and turn off the RF trap while the d.c. voltages remain constant. This transfers the $ \text{Ba}^{+}-\text{Rb}$ ensemble into the biODT. To establish reference conditions by dismissing the $ \text{Rb}$, we illuminate the ensemble with laser pulses resonant with the $ 5 ~ ^2 S_{1/2} \rightarrow  5 ~ ^2 P _{3/2}$ transition in $ ^{87} \text{Rb}$ which do not affect $ \text{Ba}^+ $. After $\tau_{int} \approx 0.5 \, \text{ms} $ with the $ \text{Ba}^+ $ in the biODT, we ramp $ P_{NIR} $ to zero and measure $ p_{opt} $ for $ \text{Ba}^+ $ in the VIS trap. The results of these reference measurements are shown as red data points (open squares) in Fig.\ref{cooling}. We recover the maximal  $ p_{opt} $ of approximately 0.8, revealing that the transfer efficiency from the hybrid trap into the biODT remains comparable to the results presented in Fig.\ref{fig:bouncers}. By fitting the data with the adapted cutoff model, we derive a temperature of $ T_{init}^{\text{Ba}^+} = 357 \pm 22 \, \mu \text{K} $, in agreement with both the result of Fig.\ref{fig:bouncers} and $ T_D^{\text{Ba}^+} $.

We now repeat this experiment but maintain the overlap of $ \text{Ba}^+ $ with the $ \text{Rb}$ atoms during the bichromatic phase. In presence of a $ \text{Rb}$ ensemble in the biODT we observe a significant increase of $ p_{opt} $ as shown in Fig.\ref{cooling} (blue circles). The corresponding temperature amounts to $ T_{symp}^{\text{Ba}^+} = 259 \pm 10 \, \mu \text{K} $. That is, we demonstrate the onset of sympathetic cooling of $ \text{Ba}^+ $, reducing the energy by $k_B \Delta T_{symp} = k_B ( - 98 \pm 24 ) ~ \mu \text{K}$ through interaction with the surrounding ensemble of $ \text{Rb}$ atoms. This result is in good agreement with our estimation of the Langevin collision rate of $  \gamma_{Lgvn} \approx 1.5 \pm 0.4 \times 10^3 \, \text{s}^{-1} $ ($ \text{Rb}$ density of $n_0 \approx 5 \times 10^{11} \, \text{cm}^{-3} $) based on the measured average atom temperature and interaction duration\,\cite{Hoeltkemeier2016}. This corresponds to, on average, one Langevin collision. Our finding indicates that the sympathetic cooling mechanism is highly efficient with a measured rate of $ R_{symp} = - 196 \pm 48 ~ \text{mK} ~ \text{s}^{-1}$, removing about a third of the ion's kinetic energy after a single interaction event.

We note that in our current parameter regime, we still face restrictions attributed to residual photo-assisted ionization of neutral rubidium and three-body recombination processes in the bichromatic trap. Shot-to-shot fluctuations of trap overlap and the resulting atomic densities currently limit the accessible interaction time to the range of $ \text{ms} $\,\cite{Kruekow2016}. This is three orders of magnitude shorter than the obtainable lifetimes in our apparatus\,\cite{Lambrecht2017}. Prolonged $ \tau_{int} $ would allow to significantly reduce atom densities, suppressing the three-body loss rate scaling as $ \gamma_{3b} \propto n_0^{2} $, whereas the density dependence of the Langevin rate is $ \gamma_{Lgvn} \propto n_0 $. In our system, approximately 10 collision events are predicted to be sufficient for reaching thermal equilibrium of $ \text{Ba}^+ $ with an ultracold $ \text{Rb}$ ensemble\,\cite{Metcalf1999}. Here, $ T^{Rb} \approx 30 ~ \mu \text{K} $ in the biODT due to conservatively chosen parameters to ensure $ n_0 < 10^{12} ~ \text{cm}^{-3}$ and correspondingly moderate $ \gamma_{3b} $. %
Note that in our NIR trap, quantum degenerate ensembles also have been achieved. With our approach the ion energy is expected to obey the Boltzmann distribution, in contrast to the non-thermal distributions observed in hybrid traps\,\cite{DeVoe2009,Meir2018,Rouse2017}. This may be advantageous for experiments in the quantum regime. Dipole trap lasers operated at optimized detunings and initialization of the ion at sub-Doppler temperatures\,\cite{Karpa2013,Meir2016} would allow to substantially decrease the optical intensity initially required to trap the $ \text{Ba}^+ $ ion, reducing the generation rates of $ \text{Rb}^+ $/$ \text{Rb}_2^+ $ ions. Similarly, using a different element, such as fermionic $ ^{6}\text{Li} $, may allow to pre-cool $ \text{Ba}^+ $ to lower energies in the hybrid trap according to theoretical predictions\,\cite{Cetina2012} or, as recently demonstrated, for the $ ^{6}\text{Li}-{}^{171}\text{Yb}^+ $ system\,\cite{Feldker2019}. With the outlined technical improvements, thermal equilibration of $ \text{Ba}^+ $ and $ \text{Rb}$ should be within reach in the current setup. Another approach to address the limitations arising from three-body recombination would be to either tune the scattering length by utilizing Feshbach resonances\,\cite{Cornish2000,Donley2008} or to employ fermionization by preparing a one-dimensional gas of $\text{Rb}$ atoms\,\cite{Goold2010}.

To summarize, we have demonstrated a method for preparing, overlapping and observing the interactions between ions and neutral atoms in an optical trap while maintaining isolation from parasitic ions. The complete absence of RF fields eliminates micromotion-induced heating, a long-standing obstacle for observing ultracold ion-atom interactions in hybrid traps\,\cite{Cetina2012}. We have further shown sympathetic cooling of $ \text{Ba}^+ $ below its initial Doppler temperature. Furthermore, our approach is not limited to extremely large ion-atom mass ratios, Rydberg atoms or homonuclear samples and may provide a generic method for establishing interactions between atoms and ion(s) in the s-wave scattering regime, even in higher dimensional crystalline structures.
\begin{acknowledgments}   
This project has received funding from the European Research Council (ERC) under the European Union’s Horizon 2020 research and innovation program (Grant No. 648330). J. S., F. T. and P. W. acknowledge support from the DFG within the GRK 2079/1 program. P. W. gratefully acknowledges financial support from the Studienstiftung des deutschen Volkes. We are indebted to V. Vuleti\'{c}, D. Leibfried, R. Moszynski, M. Tomza and O. Dulieu for many fruitful discussions.
\end{acknowledgments}   
%
%
\end{document}